# Some aspects of diffracted waves formation at RHEED on a reconstructed crystal face

A.V. Vasev[*], M.A. Putyato, V.V. Preobrazhenskii

*Institute of Semiconductor Physics, Siberian Branch of Russian Academy of Sciences, Laboratory of Physical Bases of Semiconductor Heterostructures Epitaxy, Acad. Lavrentiev Avenue, 13, 630090, Novosibirsk, Russia*

ABSTRACT

The evolution of RHEED reflexes intensity during reconstructed transitions characterizes (often implicitly) reconstructed surface state peculiarities. The approaches of a correct RHEED data interpretation, aimed at obtaining information about reconstructed transitions kinetics, are considered in the present work. In particular, the nature of RHEED reflexes formation, depending on such parameters as the average size of reconstructed domains and number of such domains per area unit, is analyzed within the kinematic approximation of the diffraction theory. This geometrical description is a convenient and effective (productive) way of analyzing reconstructed transitions mechanisms and parameters.

The transformation of the functional dependence between the measured values (RHEED reflexes intensity picture) and the degree of surface coverage by reconstruction domains, at a change of these domains average size and distribution density, is shown. This work provides the community with a useful framework for such type of theoretical studies.

*Keywords*: Reflected High-Energy Electron Diffraction; Kinematic approximation; Surface reconstructions; RHEED reflexes intensity formation

## 1. Introduction

The reflection high energy electron diffraction (RHEED) method is widely used for studying structural peculiarities of crystal surfaces. It is conditioned by the high informativity of this method that allows carrying out *in-situ* investigations of epitaxial growth processes and vacuum annealing in real time. RHEED data contain information about the composition, structure and morphology of a surface under study. Modern theoretical models [1-6] and computing powers [7, 8] enable the spatial distribution restoration for an electron wave reflected from a crystal surface in real time, with the macro (facets [9, 10], islands [11-14], terraces [12, 15-18]) and micro (roughness [19], reconstruction [20-24]) structure of this surface taken into account. Such approach (solution of the *direct* task) is of the biggest demand when monitoring the MBE growth [6, 25-29]. Besides, it is topical in the RHEED rocking curves analysis for determining the structural properties of the superstructural states under study [30-36].

At the same time, the analysis of RHEED reflexes intensity changes picture during superstructural transitions for obtaining the information about the kinetics of the processes proceeding threat [37-45] requires the solution of already a more complicated *reverse* task. It is very difficult to restore the reflected wave spatial distribution based only on the data of RHEED reflexes intensity picture, without the *a priori* known information about the composition, structure and rearrangement mechanisms of reconstructed layers.

As a consequence, there is a necessity for the development of new approaches to analyzing experimental RHEED data. This problem is most acute in investigating the processes having a complicated multistep nature. Simplest techniques (revealing exponential dependence in a measured data sequence) may lead to an erroneous interpretation of results in studying them. The activation energies obtained as a sequence of using such approach will have incorrect values. The problem of revealing the explicit (functional) dependence between the measured values (RHEED reflexes intensity) and investigated parameters of a system (e.g. degree of surface coverage by reconstruction domains) becomes topical.

---
[*] Tel.: +7 (383) 333 1967; fax: +7 (383) 333 3502.
*E-mail address*: vasev@isp.nsc.ru.

The objective of the present contribution is developing new approaches, at which the linear sizes finiteness of coherence area is taken into account, to the problem of spatial distribution restoration for an electron wave reflected from a crystal surface.

The task of finding out the dependence between the RHEED reflexes intensity and degree of surface coverage by reconstruction domains, depending on the average size of reconstruction domains and the number of these domains per area unit, is solved in the present work.

## 2. Theory

After the interaction of plane monochromatic wave $E = E_0 \cdot e^{i(k \cdot x - \omega t)}$ with a crystal, elastically scattered waves diffract with each other. But only coherent waves will diffract.

As is known, any real electron source has finite sizes. As a consequence, the electrons emitted by this source cannot be focused in an ideally parallel beam. The angular spread $\Delta\vartheta$ of this beam will determine the size of the area where electron waves will be coherent. The length of coherence area in the direction perpendicular to the beam (coherence area width) equals $L_\perp = \lambda / \Delta\vartheta$. In the parallel direction the coherence length is $L_\parallel = \dfrac{\lambda}{\Delta\vartheta \cdot \sin\vartheta}$. In other words, a coherence area ellipse, oriented along the direction of wave vector projection $xx'$, with a big semiaxis equal to $L_\parallel/2$ and a small semiaxis equal to $L_\perp/2$, forms on the investigated surface.

When considering the crystal, let us use its presentation as a primitive 3D lattice, each site of which is compared to an elementary cell and basis. Generally, let the primitive lattice be characterized by the lowest (triclinic) syngony ($a_1 \neq a_2 \neq a_3$ and $\phi_1 \neq \phi_2 \neq \phi_3 \neq 90°$).

It is necessary to ascribe the scattered wave $E_{S,l}$, in the formation of which all primitive lattice sites - being in the coherence area with the center in this site $l$ - will participate, to each site of crystal $l$:

$$E_{S,l} = E \frac{1}{N_{uc}} \sum_j^{N_{uc}} S_j \cdot e^{i(s \cdot r_j)}, \qquad (2.1)$$

where $s = k_s - k_0$ – scattering vector; $k_0$ and $k_s$ – wave vectors of incident and scattered waves, respectively; $r_j = v_{j,1} a_1 + v_{j,2} a_2 + v_{j,3} a_3$ – radius-vector drawn from the beginning of coordinates to site $j$ ($v_{j,1}$, $v_{j,2}$ and $v_{j,3}$ – integers). The summing in expression (2.1) is done over all sites in the coherence area ($N_{uc}$ – full number of such sites (*unit cells*)).

If the *basis* of such crystal elementary cell consists of $N_b$ scattering centers, then, factor $S_j = \sum_g^{N_b} F_g \cdot e^{i(s \cdot \rho_{j,g})}$ is a structural factor, where the nucleus position of the g$^{th}$ scattering center, relative to site $j$, is determined by vector $\rho_{j,g}$. The atomic scattering factor (or formfactor) is set by the expression $F_g = \dfrac{8\pi^2 m}{h^2} \dfrac{e^2}{s^2}(Z_g - F_{e^-})$, where $Z_g$ – number of electrons in the atom of scattering center $g$. Here $F_{e^-}$ – the ratio of the amplitude scattered by the whole atom and the amplitude scattered by one electron.

Thus, the reflex intensity of RHEED picture can be written as

$$I = \frac{e^{-2MT}}{R^2} \sum_l^V \left( E^*_{S,l} E_{S,l} \right). \qquad (2.2)$$

Here, the summing is already over all surface sites $V$, which are in the area formed by an incident electron beam and it has microscopic dimensions. This area is a geometrical place of points, which satisfies the condition of crossing the Laue sphere with a luminescent screen plane (in experimental device geometry). Factor $e^{-2MT}$ – Debye–Waller factor – is a dimensionless value that characterizes the influence of crystal lattice thermal vibrations on radiation elastic scattering processes in a crystal [46].

As wave $E_{S,l}$ is the superposition of the waves scattered over a crystal bulk and surface $E_{S,l} = E_{SB,l} + E_{SS,l}$, then it is possible to write down:

$$E^*_{S,l} E_{S,l} = \left( E^*_{SB,l} + E^*_{SS,l} \right) \cdot \left( E_{SB,l} + E_{SS,l} \right) =$$
$$= E^*_{SB,l} E_{SB,l} + E^*_{SS,l} E_{SS,l} + E^*_{SB,l} E_{SS,l} + E^*_{SS,l} E_{SB,l}.$$
$$(2.3)$$

The contribution of the bulk component in expression (2.3) remains unchanged in the structural rearrangements process. As a consequence, the dependence of RHEED-picture reflexes intensity will have the components linear and quadratic on $E_{SS,l}$.

A 2D lattice with also the lowest syngony ($b_1 \neq b_2$ and $\varphi \neq 90°$) will be observed on an arbitrary low-index crystal face. Consider the superstructural transition $\alpha \to \beta$, characterized by the absence of intermediate states and fulfilling condition $Q_\alpha + Q_\beta = 1$, on this face. Here, $Q_\alpha$ and $Q_\beta$ – degrees of surface coverage by states $\alpha$ and $\beta$, respectively. The elementary cells of states $\alpha$ and $\beta$ are characterized by symmetries $n \times m$ and $q \times p$, respectively.

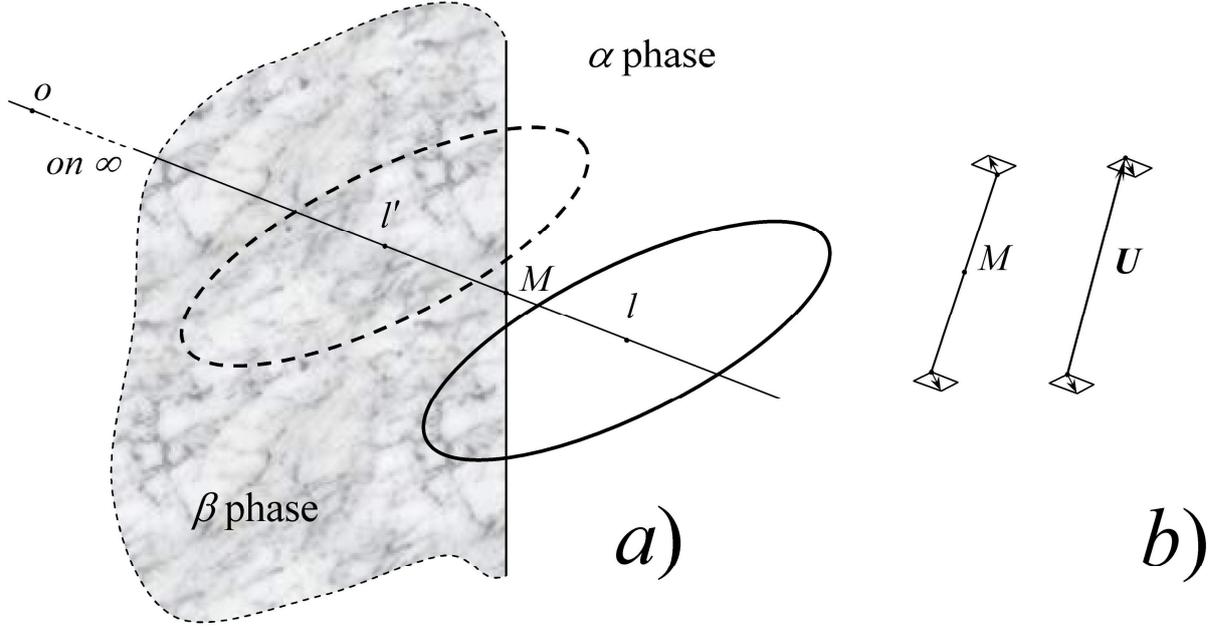

**FIGURE 1** *a*) Coherence area on the border between the domains half-planes; *b*) the influence of inversion transformations and parallel transfer on the elementary cell orientation.

Let us analyze different variants of $E_{SS,l}$ formation.

The first two of them are connected with the situations when the coherence area is wholly placed either in the domain of state $\alpha$, or in the domain of state $\beta$. Thereat, we obtain $E_{SS,l} \equiv E_{SS,l}(\alpha)$ or $E_{SS,l} \equiv E_{SS,l}(\beta)$, respectively.

The next variant is realized when the coherence area crosses the border between the domains of states $\alpha$ and $\beta$, the domains dimensions being much larger than those of the coherence area (on the scale of coherence area, the domains can be considered as half-planes divided by a straight borderline.

Let us consider the coherence area with the center in site $l$ (see Fig. 1 a). Let the part of it, that is occupied by state $\alpha$, have $N_\alpha$ elementary cell sites. Then there will be $N_\beta$ sites (here $N = nm \cdot N_\alpha + qp \cdot N_\beta$) for state $\beta$. Taking into account (2.1), it is possible to write for $E_{SS,l}$ the following:

$$E_{SS,l} = E \cdot \frac{1}{N}\left( S(\alpha)\sum_j^{N_\alpha} e^{i(s \cdot r_j)} + S(\beta)\sum_j^{N_\beta} e^{i(s \cdot r_j)} \right).$$
(2.4)

Present the elementary cell structural factors of states $\alpha$ and $\beta$ as

$$S(\alpha) = \frac{S(\alpha)}{nm}\sum_{g=0}^{n}\sum_{h=0}^{m} e^{i(s \cdot \rho_{j,gh})} = S'(\alpha)\sum_{g=0}^{n}\sum_{h=0}^{m} e^{i(s \cdot \rho_{j,gh})},$$
(2.5)

and

$$S(\beta) = \frac{S(\beta)}{qp}\sum_{g=0}^{q}\sum_{h=0}^{p} e^{i(s \cdot \rho_{j,gh})} = S'(\beta)\sum_{g=0}^{q}\sum_{h=0}^{p} e^{i(s \cdot \rho_{j,gh})},$$
(2.6)

where $\rho_{j,gh} = g\boldsymbol{b}_1 + h\boldsymbol{b}_2$.

For the $s$ vectors, corresponding to Bragg reflexes, ratio $e^{i(s \cdot \rho_{j,gh})} = 1$ is fulfilled, as it satisfies the Laue condition (further, we will use this fact quite often in our speculations).

Considering (2.5) and (2.6), expression (2.4) can be rewritten as

$$E_{SS,l} = \frac{E}{N}\left( S'(\alpha)\sum_j^{N'_\alpha} e^{i(s \cdot R_j)} + S'(\beta)\sum_j^{N-N'_\alpha} e^{i(s \cdot R_j)} \right).$$
(2.7)

Here, $\boldsymbol{R} \equiv \boldsymbol{r} + \boldsymbol{\rho}$ and $N'_\alpha \equiv nm \cdot N_\alpha$.

Let us draw a straight line through site $l$ and the domain center with state $\beta$ (in this case, it is point $o$ being at an infinite distance from the border). As point $o$ is infinitely distant, there is some arbitrariness in the choice of its location. Let us demand that straight line *lo* be located along one of the main crystallographic directions of crystal face; for convenience, choose the direction which is the closest to the normal towards the border between the domains.

We will move along straight line *lo* towards point *o* and realize, thereat, a parallel transfer of the coherence area. On the structuring, there is such position of the transferred area (point $l'$ on straight line *lo*) when the part of it occupied by state $\alpha$, in its area, is equivalent to the part of coherence area with the center in site *l*, which is occupied by state $\beta$ (see Fig. 1 a). It is easy to note that they are identical due to the inversion relative to point *M*.

The statement about point $l'$ can be more substantiated. For this purpose, it is necessary to note that the consideration is made in the discrete system – at the crystal border that consists of atoms. Thus, the straight borderline between the domains, in practice, is a broken straight line stretching through the 2D lattice sites located maximally close to the border. As a sequence of it, point $l'$ is a lattice site.

The next important assertion takes into account the fact that line *lo* is located along one of the main crystallographic directions of the crystal face. The elementary cells of states $\alpha$ and $\beta$ will be co-oriented along line *lo* by one of their sides. Such cells, relative to point *M*, will preserve their edges orientation (Fig. 1 b). It means that the inversion related to point M is the method of single-correspondence transfer of a dense cell package *scheme* at the transition from one to the other (and vice versa). The cells proper, thereat, are moved according to the circuit positions with the method of *parallel* transfer at the translation vector $U = u_1 b_1 + u_2 b_2$; here $u_1$ and $u_2$ are integers. The result of the transfer procedure is presented in Fig. 2. For site $l'$ we have:

$$E_{SS,l'} = \frac{E}{N}\left( S(\alpha) \sum_j^{N_{all,\alpha}-N_\alpha} e^{i(s \cdot r'_j)} + S(\beta) \sum_j^{N_{all,\beta}-N_\beta} e^{i(s \cdot r'_j)} \right),$$
(2.8)

here $N_{all,\alpha} \equiv N/nm$ и $N_{all,\beta} \equiv N/qp$.

Taking into account (2.5) and (2.6), expression (2.8) can be rewritten as

$$E_{SS,l'} = \frac{E}{N}\left( S'(\alpha) \sum_j^{N-N'_\alpha} e^{i(s \cdot R_j)} + S'(\beta) \sum_j^{N'_\alpha} e^{i(s \cdot R'_j)} \right) =$$

$$= \frac{E}{N}\left( S'(\alpha) \sum_j^{N-N'_\alpha} e^{i(s \cdot R_j)} e^{-i(s \cdot U_j)} + S'(\beta) \sum_j^{N'_\alpha} e^{i(s \cdot R'_j)} \right) =$$

$$= \frac{E}{N}\left( S'(\alpha) \sum_j^{N-N'_\alpha} e^{i(s \cdot R_j)} + S'(\beta) \sum_j^{N'_\alpha} e^{i(s \cdot R'_j)} \right).$$
(2.9)

Analogously, we will have for (2.7):

$$E_{SS,l} = \frac{E}{N}\left( S'(\alpha) \sum_j^{N'_\alpha} e^{i(s \cdot R_j)} + \right.$$

$$\left. + S'(\beta) \sum_j^{N-N'_\alpha} e^{i(s \cdot R_j)} e^{i(s \cdot U_j)} \right) =$$

$$= \frac{E}{N}\left( S'(\alpha) \sum_j^{N'_\alpha} e^{i(s \cdot R_j)} + S'(\beta) \sum_j^{N-N'_\alpha} e^{i(s \cdot R'_j)} \right),$$
(2.10)

here $R' = R + U$ and $e^{\pm i(s \cdot U)} = 1$.

It is convenient to group the quadratic and linear members (see expression (2.3)) pairwise:

$$E^*_{SS,l} E_{SS,l} + E^*_{SS,l'} E_{SS,l'} =$$

$$\frac{E^* E}{N^2}\left[ \left( S'(\alpha) \sum_j^{N} e^{i(s \cdot R_j)} \right)^* \left( S'(\alpha) \sum_j^{N} e^{i(s \cdot R_j)} \right) + \right.$$

$$+ \left( S'(\beta) \sum_j^{N} e^{i(s \cdot R'_j)} \right)^* \left( S'(\beta) \sum_j^{N} e^{i(s \cdot R'_j)} \right) -$$

$$- \left( S'(\alpha) - S'(\beta) \right)^* \left( S'(\alpha) - S'(\beta) \right) \cdot$$

$$\left. \cdot \left( \sum_j^{N'_\alpha} e^{-i(s \cdot R_j)} \cdot \sum_j^{N-N'_\alpha} e^{i(s \cdot R_j)} + \sum_j^{N'_\alpha} e^{i(s \cdot R_j)} \cdot \sum_j^{N-N'_\alpha} e^{-i(s \cdot R_j)} \right) \right] =$$

$$= E^*_{SS,l}(\alpha) E_{SS,l}(\alpha) + E^*_{SS,l'}(\beta) E_{SS,l'}(\beta) -$$

$$- E^* E \cdot \Delta S'^* \Delta S' \cdot \eta .$$
(2.11)

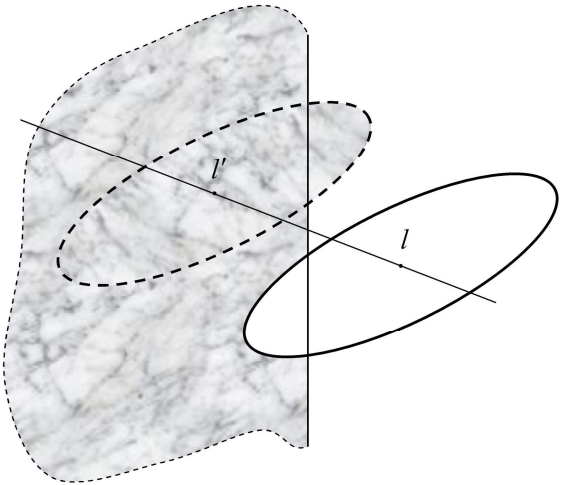

**FIGURE 2** Elementary cells redistribution inside pair coherence areas with centers *l* and $l'$ (to a semi-infinite domain).

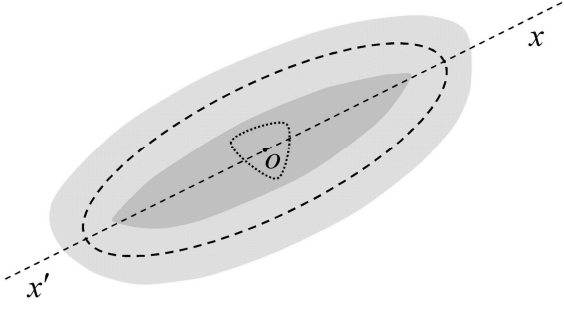

**FIGURE 3** Diagram of conditions for realizing three possible variants of coherence area location related to the domain.

Here $\Delta S' = S'(\alpha) - S'(\beta)$, and, for the sum in the latter brackets

$$\eta \equiv \frac{1}{N^2}\left(\sum_j^{N'_\alpha} e^{-i(s\cdot R_j)} \cdot \sum_j^{N-N'_\alpha} e^{i(s\cdot R_j)} + \right.$$
$$\left. +\sum_j^{N'_\alpha} e^{i(s\cdot R_j)} \cdot \sum_j^{N-N'_\alpha} e^{-i(s\cdot R_j)} \right),$$

we have: $\eta \in \mathbb{R}$ and $0 \leq \eta \leq 1$.

At fulfilling condition $\Delta S'^* \Delta S' \cdot \eta \ll 1$, just is $E_{SS,l}^* E_{SS,l} + E_{SS,l'}^* E_{SS,l'} \approx E_{SS,l}^*(\alpha)E_{SS,l}(\alpha) + E_{SS,l'}^*(\beta)E_{SS,l'}(\beta)$ for any moment of the spread of coherence area through the domain border.

For linear members, the pairwise grouping leads to expression $E_{SS,l} + E_{SS,l'} =$

$$= \frac{E}{N}\left(S(\alpha)\sum_j^{N_{all,\alpha}} e^{i(s\cdot R_j)} + S(\beta)\sum_j^{N_{all,\beta}} e^{i(s\cdot R'_j)}\right) =$$
$$= E_{SS,l}(\alpha) + E_{SS,l'}(\beta). \quad (2.12)$$

The physical essence of this statement can also be easily seen in Fig. 2. As for (2.11), expression (2.12) is also just for any moment of coherence area spreading through the domain border.

We will remind of that the above-considered variant is just for the situation when the domain sizes of states $\alpha$ and $\beta$ are bigger than the coherence area sizes. Let it be not so, and the domain size, e.g. of state $\beta$, is smaller than the coherence area sizes. In this case, one of three variants can be realized. The first one (trivial and already considered earlier) is when the coherence area and the domain do not cross. The second – when part of the domain area is located in the coherence area. The third one – when all domain is wholly located inside the coherence area.

A diagram in Fig. 3 is presented to illustrate the conditions of realizing each of three variants. If the center of the coherence area is located inside the area in dark-grey color, then this is the third variant. If the center is inside the light-grey area, then we deal with the second variant. And, finally, if the center is in the white area, the first variant is realized. It is worth noting that, at the location of the center of coherence area in the light-grey area inside/outside the coherence area contour (dashed ellipse), then more/less than a half of the domain area will be located in the coherence area. The dark- and light-grey areas shapes, respectively, are also changed at a change of the domain shape, the dashed border position being unchanged. Consider two arbitrary sites $l_A$ and $l_B$ located in the dark-grey area (Fig. 4). These sites are distant from each other at the translation vector $u_{AB}$ belonging to a set of $U$ vectors. Having indicated the number of elementary cells in the domain with state $\beta$ as $N_{\beta 0} \equiv N'_{\beta 0}/qp$, it is possible to put down for $E_{SS,l_A}$ and $E_{SS,l_B}$:

$$E_{SS,l_A} = \frac{E}{N}\left(S'(\alpha)\sum_j^{N-N'_{\beta 0}} e^{i(s\cdot R_{A,j})} + S'(\beta)\sum_j^{N'_{\beta 0}} e^{i(s\cdot R_{A,j})}\right) =$$
$$= \frac{E}{N}\left(S'(\alpha)\sum_j^{N-N'_{\beta 0}} e^{i(s\cdot R_{B,j})}e^{-i(s\cdot u_{AB})} + S'(\beta)\sum_j^{N'_{\beta 0}} e^{i(s\cdot R_{B,j})}\right) =$$
$$= \frac{E}{N}\left(S'(\alpha)\sum_j^{N-N'_{\beta 0}} e^{i(s\cdot R_{B,j})} + S'(\beta)\sum_j^{N'_{\beta 0}} e^{i(s\cdot R_{B,j})}\right) \equiv E_{SS,l_B}.$$
(2.13)

Here $e^{-i(s\cdot u_{AB})} = 1$, $R_A \equiv R_B$ for sites inside the domain, and $R_A = R_B + u_{AB}$ for all the rest sites inside the coherence area.

Thus, the diffracted wave intensities, corresponding to all sites of the dark-grey area, will be the same and equal to expression (2.13). Expression (2.13) is also the definition of a new homogeneous phase $\beta^*$, for which just is

$$E_{SS,l}(\beta^*) \equiv$$

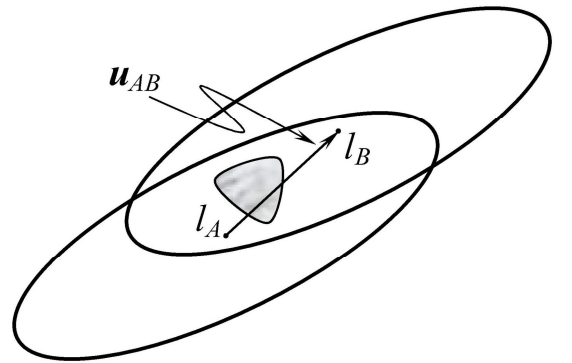

**FIGURE 4** Small domain in the coherence area.

$$\equiv \frac{E}{N}\left(S'(\alpha)\sum_{j}^{N-N'_{\beta0}}e^{i(s\cdot R_{B,j})}+S'(\beta)\sum_{j}^{N'_{\beta0}}e^{i(s\cdot R_{B,j})}\right). \quad (2.14)$$

Now consider the situation when the center of coherence area (site $l$) is in the light-grey area (second variant, Fig. 4). As earlier, let us draw a straight line $lo$ through site $l$ and the domain center with state $\beta$ (point $o$). It is convenient to choose the position of the site closest to the domain mass center as point $o$.

Note that the procedure of building a pair coherence area with the center in point $l'$ requires carrying out the obligatory condition, i.e. the unambiguity of reflecting a set of $l$ points into a set of $l'$ points related to site $o$. This condition is fulfilled automatically for domains with a convex border topology. For domains with a concave border topology, the task is solved by breaking a domain into convex components, which complicates calculations, but it does not change the gist of the approach. Thus, further on, it will run only about domains with a convex border topology. Just as it was earlier, we will move along line $lo$ in the direction from site $l$ towards site $o$ (and farther) and, therein, realize a parallel transfer of the coherence area. There is such position of the transferred area center (point $l'$ on line $lo$), at which a domain part, that is not in the coherence area with the center in site $l$ (Fig. 5 a), will be in this transferred area. A more detailed scheme of dividing the domain by borders can be seen in Fig. 5 b.

Let the borders of coherence areas with their centers in points $l$ and $l'$ cross in points $z$ and $z'$. Designate the crossing point of lines $zz'$ and $lo$ as $M$. As it was mentioned earlier, line $zz'$ is really a broken line stretching through 2D lattice sites.

Hence, point $l'$, being the inversion of site $l$ related to point $M$, is also a site.

In Fig. 5 b, on the structuring, the total area of two sectors (dark-grey) is equal to the total area of the lens that consists of two segments (light-grey). As a sequence, the number of elementary cells for the state $\beta$ in the dark-grey areas is equal to the number of elementary states of the same state in the light-grey area. These cells are equivalent to each other with the accuracy to a parallel translation to the vector belonging to a set of $U$ vectors.

Realize the transformation of the parallel transfer for the domain part that was earlier located in area with its center in site $l'$ into the coherence area with its center in site $l$ (Fig. 6). The transfer is realized to vector $l'l$. It is obvious that this vector belongs to a set of $U$ vectors. After the transfer, let us fill in the empty light-grey lens area with the elementary cells from the dark-grey sectors. As a result, all the domain with state $\beta$ will be inside the coherence area with center $l$, and the coherence area with center $l'$ will have only the elementary cells of state $\alpha$.

Then, for $E_{SS,l}$ and $E_{SS,l'}$, we can write:

$$E_{SS,l}=$$
$$=\frac{E}{N}\left(S'(\alpha)\sum_{j}^{N-N'_{\beta}}e^{i(s\cdot R_{j})}+S'(\beta)\sum_{j}^{N'_{\beta}}e^{i(s\cdot R_{j})}\right)=$$
$$=\frac{E}{N}\left(S'(\alpha)\sum_{j}^{N-N'_{\beta0}}e^{i(s\cdot R_{j})}+S'(\beta)\sum_{j}^{N'_{\beta0}}e^{i(s\cdot R_{j})}+\right.$$
$$\left.+(S'(\alpha)-S'(\beta))\sum_{j}^{N'_{\beta0}-N'_{\beta}}e^{i(s\cdot R_{j})}\right), \quad (2.15)$$

и

$$E_{SS,l'}=$$

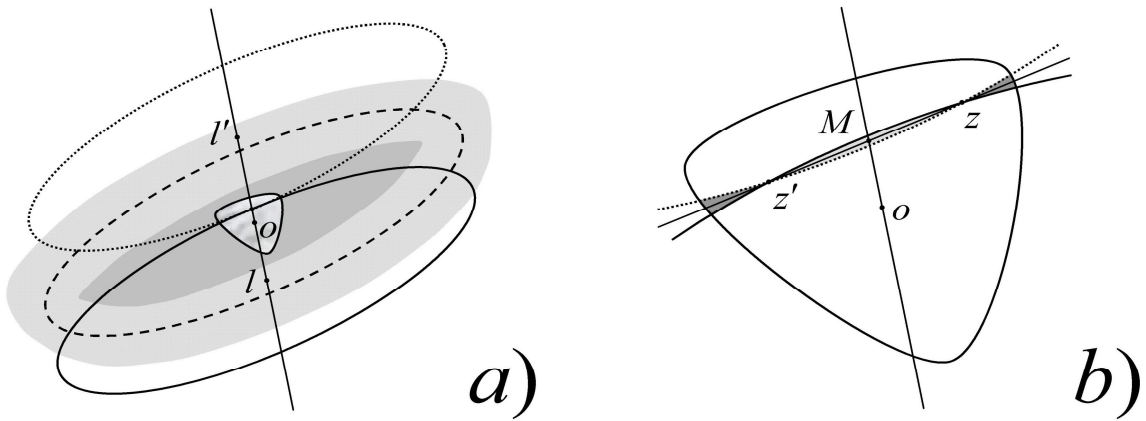

**FIGURE 5** *a*) Mutual position of coherence areas with centers $l$ and $l'$ when they share a small-sized domain; *b*) detailed scheme of dividing the domain by the borders of coherence areas.

$$= \frac{E}{N}\left( S'(\alpha) \sum_{j}^{N-(N'_{\beta 0}-N'_{\beta})} e^{i(s\cdot R'_j)} + S'(\beta) \sum_{j}^{N'_{\beta 0}-N'_{\beta}} e^{i(s\cdot R'_j)} \right) =$$

$$= \frac{E}{N}\left( S'(\alpha)\sum_{j}^{N} e^{i(s\cdot R'_j)} - \right.$$

$$\left. -(S'(\alpha)-S'(\beta))\sum_{j}^{N'_{\beta 0}-N'_{\beta}} e^{i(s\cdot R'_j)}e^{-i(s\cdot u_{ll'})} \right) =$$

$$= \frac{E}{N}\left( S'(\alpha)\sum_{j}^{N} e^{i(s\cdot R_j)} - \right.$$

$$\left. -(S'(\alpha)-S'(\beta))\sum_{j}^{N'_{\beta 0}-N'_{\beta}} e^{i(s\cdot R_j)} \right). \quad (2.16)$$

Here $R' = R + u_{ll'}$ and $e^{-i(s\cdot u_{ll'})} = 1$.

Thus, for $E^*_{SS,l}E_{SS,l} + E^*_{SS,l'}E_{SS,l'}$, we have:

$$\frac{E^*E}{N^2}\left[\left( S'(\alpha)\sum_{j}^{N-N'_{\beta 0}} e^{i(s\cdot R_j)} + S'(\beta)\sum_{j}^{N'_{\beta 0}} e^{i(s\cdot R_j)} \right)^* \cdot \right.$$

$$\cdot\left( S'(\alpha)\sum_{j}^{N-N'_{\beta 0}} e^{i(s\cdot R_j)} + S'(\beta)\sum_{j}^{N'_{\beta 0}} e^{i(s\cdot R_j)} \right) +$$

$$+\left( S'(\alpha)\sum_{j}^{N} e^{i(s\cdot R'_j)} \right)^*\left( S'(\alpha)\sum_{j}^{N} e^{i(s\cdot R'_j)} \right) -$$

$$-(S'(\alpha)-S'(\beta))^*(S'(\alpha)-S'(\beta))\cdot$$

$$\left. \cdot\left( \sum_{j}^{N'_{\beta}} e^{-i(s\cdot R_j)}\cdot\sum_{j}^{N-N'_{\beta}} e^{i(s\cdot R_j)} + \sum_{j}^{N'_{\beta}} e^{i(s\cdot R_j)}\cdot\sum_{j}^{N-N'_{\beta}} e^{-i(s\cdot R_j)} \right) \right] =$$

$$= E^*_{SS,l}(\beta^*)E_{SS,l}(\beta^*) + E^*_{SS,l'}(\alpha)E_{SS,l'}(\alpha) -$$

$$- E^*E\cdot\Delta S'^*\Delta S'\cdot\eta. \quad (2.17)$$

According to expressions (2.13) and (2.17), the wave scattering process on the small ($N'_{\beta 0} < N$) domain with structural state $\beta$ can be presented as a scattering process of the same wave, but on a domain of bigger sizes (equal to coherence area sizes), filled in by a new efficient structural state $\beta^*$. The scattering properties of this state are determined by expression (2.14).

In other words, the diffracted wave intensities will be the same, being equal in their values to $E_{SS,l}(\beta^*)$ from expression (2.14) for all sites located inside the dashed border in the diagram of Fig. 3. For the sites outside the dashed borderline, the intensity will equal $E_{SS,l}(\alpha)$. It also follows from Fig. 6.

Thus, having replaced small domains by big domains (the centers position should coincide), we are, automatically, in the conditions of the earlier-described variant. However, there will be one important distinction observed herein. The domains will lose their "individuality in sizes". All domains with new states will have the same coherent area sizes.

We would like to stress out one important moment. At small domain sizes, the situation, when the ratio of domain area to the square of coherence area $\mu$ becomes a small value, is realized:

$$S'(\alpha) \sum_{j}^{N-N'_{\beta 0}} e^{i(s\cdot R_j)} + S'(\beta)\sum_{j}^{N'_{\beta 0}} e^{i(s\cdot R_j)} =$$

$$= S'(\alpha)\sum_{j}^{N} e^{i(s\cdot R_j)} - (S'(\alpha)-S'(\beta))\sum_{j}^{N'_{\beta 0}} e^{i(s\cdot R_j)} =$$

$$= \left(1 - \frac{\Delta S'}{S'(\alpha)}\cdot\mu\right)\cdot S'(\alpha)\sum_{j}^{N} e^{i(s\cdot R_j)}, \quad (2.18)$$

where $0 < \mu \ll 1$.

Thus,

$$E^*_{SS,l}(\beta^*)E_{SS,l}(\beta^*) + E^*_{SS,l'}(\alpha)E_{SS,l'}(\alpha) \approx$$

$$\approx \left(2 - \frac{\Delta S'}{S'(\alpha)}\cdot\mu - \frac{\Delta S'^*}{S'^*(\alpha)}\cdot\mu^*\right)\cdot E^*_{SS,l'}(\alpha)E_{SS,l'}(\alpha) =$$

$$= E^*_{SS,l'}(\alpha)E_{SS,l}(\beta^*) + E_{SS,l'}(\alpha)E^*_{SS,l}(\beta^*). \quad (2.19)$$

As a sequence, the scattered wave intensity will be determined by linear components. In other words, the scattered wave intensity dependence on the degree of surface coverage by domains will be *linear*.

Obviously, the analogous speculations are just

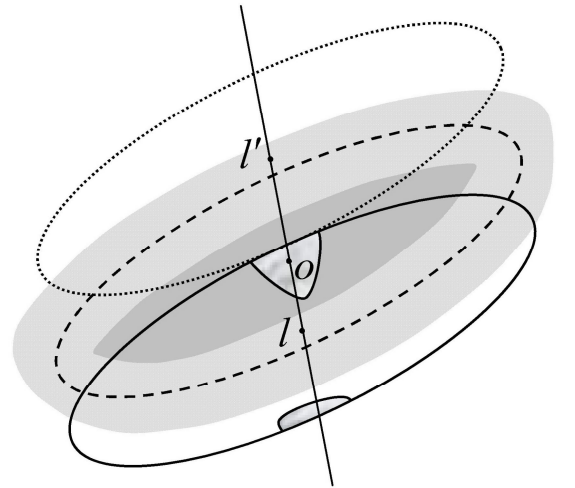

**FIGURE 6** Elementary cells redistribution within the pair coherence areas with centers $l$ and $l'$ (for a small domain).

also for small domains with state $\alpha$.

So, if the number of domains per area unit does not exceed $1/S_{coh}$ ($S_{coh}$ – square of coherence area) for a developing state, then the reflexes intensity of RHEED picture is the function linear on $Q_\alpha$:

$$I = Q_\alpha \cdot I(\alpha) + (1 - Q_\alpha) \cdot I(\beta). \quad (2.20)$$

On the other hand, if a small domains number per unit area begins to exceed $1/S_{coh}$, then the scattered wave intensity stops to depend on the scattering site position on the surface. Ratio $N_\alpha/N \to const$ will be fulfilled for any surface point being in the locus. At carrying out this condition, the reflected wave intensity dependence on the degree of surface coverage by domains of phases $\alpha$ and $\beta$ stops to be linear. It can be explicitly established with the model first proposed by C.S. Lent and P.I. Cohen in 1984 [15] and then extended by many other authors [47, 48]. According to this model, the diffracted wave intensity can be schematically presented as

$$\left\{ \sum_G \delta(s_x - G) \right\} * \{\text{Instr. Response}\} *$$
$$* \left\{ C_{order} \cdot \delta(s_x) + C_{diffus} \cdot (\text{Broadening}) \right\}, \quad (2.21)$$

i.e. as a folding over all $G$ sites of the reverse crystal lattice, realized for the instrumental function (characterizes device sensitivity [49-51]) and reflex profile shape functions). According to this expression, the diffraction reflex profile in the RHEED picture is a sum of narrow and broad peaks. The δ-function peak intensity characterizes a distant order in the domains with phases $\alpha$ and $\beta$. The surface disordering degree (determined by the presence of domain borders, also a possibility of their non-cophase arrangement) is characterized by the broad peak intensity. The broad peak shape is determined by the function of domains distribution over sizes [52]. For dimensionality models $1 + 1$ (normal + lateral direction) [47, 48], the broad peak has the Lorentz shape. Within this model, premultipliers $C_{order}$ and $C_{diffus}$ can be written in more detail: $C_{order} = Q_\alpha^2 \cdot f_\alpha^2 + (1 - Q_\alpha)^2 \cdot f_\beta^2 +$

$$+ 2Q_\alpha (1 - Q_\alpha) \cdot f_\alpha f_\beta \cdot \cos(s_z c + \phi_\beta - \phi_\alpha), \quad (2.22)$$

$$C_{diffus} = Q_\alpha (1 - Q_\alpha) \cdot$$
$$\cdot \left( f_\alpha^2 + f_\beta^2 - 2 f_\alpha f_\beta \cdot \cos(s_z c + \phi_\beta - \phi_\alpha) \right). \quad (2.23)$$

Here, $f_\alpha$ and $f_\beta$ – structural factors for phases $\alpha$ and $\beta$, respectively. Also $s_z c$ and $\phi_\beta - \phi_\alpha$ – the phase shifts caused by the difference of heights $c$ and the scattering for phases $\alpha$ and $\beta$, respectively. Thus, the reflected wave intensity dependence on the degree of surface coverage by domains is of the *mixed* kind (simultaneously quadratic and linear).

An interesting case of the loss of "individuality in sizes" is the limit situation when the size of the forming state domains tends to the size of scattering center. This variant corresponds to the *islandless* structural transition. During such transition, the all volume of structural state $\alpha$ is considered as one domain, and a transition from state $\alpha$ into state $\beta$ is probable for any scattering center. A typical example of such transition is the desorption-induced $order \to disorder$-type transition. This transition corresponds to the model description of the influence of small vacancy complexes on the diffracted wave intensity [53].

According to the approach proposed by J.M. Cowley, the diffracted wave intensity, in this case, can be presented as:

$$I = Q_\alpha^2 \cdot f_\alpha^2 \cdot \frac{1}{N} \sum_i \sum_j \exp\left(2\pi i \cdot u (r_i - r_j)\right) +$$
$$+ (1 - Q_\alpha) \cdot Q_\alpha \cdot f_\alpha^2, \quad (2.24)$$

here $Q_\alpha$ – fraction of the surface part which is covered by the defectless elementary cells of reconstruction $\alpha$. The first summand in expression (2.24) corresponds to the diffracted energy distribution for a structurally perfect surface with reconstruction $\alpha$, but with a $Q_\alpha$ decreased structural factor $f_\beta$. As for the second summand, this corresponds to the diffracted energy distribution for $(1 - Q_\alpha) \cdot N$ *isolated* defectless elementary cells (they may be neglected at small $1 - Q_\alpha$). In this case, the *quadratic* functional dependence is realized.

Thus, depending on the conditions of realizing a structural transition, it is possible to observe different types of the functional dependence of RHEED reflexes intensity and the degree of surface coverage by reconstructed domains, and they are linear, mixed, linear-quadratic and, finally, purely quadratic.

It is important to note that the RHEED rocking curves additivity is an analogue of fulfilling equation (2.20). The additivity here is understood as the fulfilment of equality

$$I_\vartheta(\alpha') = Q_\alpha \cdot I_\vartheta(\alpha) + (1 - Q_\alpha) \cdot I_\vartheta(\beta), \quad (2.25)$$

where $I_\vartheta(\alpha)$ and $I_\vartheta(\beta)$ – RHEED rocking curves for structural states $\alpha$ and $\beta$, respectively, and $I_9(\alpha')$ – RHEED state rocking curve $\alpha' = Q_\alpha \cdot \alpha + (1 - Q_\alpha) \cdot \beta$ at any transition moment. Here, $I_9$ can be considered as an integral operator.

As a sequence, carrying out the condition of additivity (2.25) can be considered as a criterion for realizing the linear functional dependence. This criterion can be most useful in the implementation of the RHEED investigations in superstructural transitions kinetics on a reconstructed crystal surface.

To confirm the fairness of our theoretical speculations, one can point out the results of the experimental work by K. Shimada *et al.* [54]. It is shown in it that RHEED can be used as a method of quantitative determining the degree of Si(111) surface coverage by domains with reconstruction (7×7). Besides, a successful use of the RHEED method for the quantitative determination of the degree of surface coverage by reconstructed domains is illustrated in the publications by S. Hasegawa *et al.* [55] and [56]. In these contributions the authors studied the change of superstructural surface Si(111) state depending on the degree of its coverage by metals Au, Al и In.

## 3. Conclusion

The nature of RHEED-picture reflexes formation with co-existing superstructural states was studied within the kinematic approach. The linear sizes finiteness of scattered electron coherence area was first taken into account at such consideration.

It is shown that domains geometrical parameters (average size and their number per area unit) determine the type of the functional dependence between the RHEED-picture reflexes intensity and the degree of surface coverage by domains with superstructural states. In particular, at fulfilling condition $1/S_{dom} < 1/S_{coh}$, this functional dependence is *linear*. If condition $1/S_{dom} > 1/S_{coh}$ is fulfilled, then the *mixed* (simultaneously quadratic and linear) functional dependence is realized. And, finally, in the limit case of $S_{dom} \to 0$ (at condition $1/S_{dom} \gg 1/S_{coh}$) it is possible to observe the *quadratic* functional dependence. Here, $S_{dom}$ and $S_{coh}$ – the average area of reconstructed domain and the coherence ellipse area, and $1/S_{dom}$ and $1/S_{coh}$ – the number of domains and coherence areas per area unit, respectively.

It is shown that the fulfillment of the RHEED rocking curves additivity condition is a trait of realizing the linear dependence between the specular-beam intensity and the degree of surface coverage by domains.

## Acknowledgement
This work was supported by Russian Science Foundation, grant № 16-12-00023.